\documentclass[pra, showpacs, twocolumn]{revtex4}
\usepackage{graphicx,xcolor}
\usepackage{amsmath, bbm}
\usepackage{hyperref}
\usepackage{epstopdf}

\renewcommand{\Im}{\mbox{Im }}
\renewcommand{\Re}{\mbox{Re }}

\renewcommand{\imath}[0]{\mathrm{i}}

\newcommand{\EXP}[1]{e^{#1}}
\newcommand{\unit}[1]{\mathrm{#1}}
\renewcommand{\vec}[1]{\boldsymbol{#1}}

\begin{document}

\title{Dynamical Casimir-Polder interaction between an atom 
and surface plasmons}
 \author{Harald R. Haakh$^{1,2}$}%
 \email{harald.haakh@mpl.mpg.de }
  \author{Carsten Henkel$^1$}
 \affiliation{
$^1$Institut f\"{u}r Physik und Astronomie, Universit\"{a}t Potsdam, Karl-Liebknecht-Stra{\ss}e 24/25, D-14476 Potsdam, Germany\\
$^2$Max Planck Institute for the Science of Light, G\"unther-Scharowski-Stra{\ss}e 1/24, D-91058 Erlangen, Germany
}
 \author{Salvatore Spagnolo, Lucia Rizzuto, and Roberto Passante}
 \affiliation{
Dipartimento di Fisica e Chimica, Universit\`{a} degli Studi di Palermo and CNISM,
 Via Archirafi 36, I-90123 Palermo, Italy}

\date{26 Nov 2013}

\begin{abstract}
We investigate the time-dependent Casimir-Polder potential of a polarizable two-level atom placed near a surface of arbitrary material, after a sudden change in the parameters of the system.
Different initial conditions are taken into account. For an initially bare ground-state atom, the time-dependent Casimir-Polder 
energy reveals how the atom is ``being dressed'' by virtual, matter-assisted photons.
We also study the transient behavior of the Casimir-Polder interaction 
between the atom and the surface starting from a partially dressed state, after an externally induced 
change in the atomic level structure or transition dipoles.
The Heisenberg equations are solved through an iterative technique for both atomic and field 
 operators in the medium-assisted electromagnetic field quantization scheme.
We analyze in particular how the time evolution of the interaction energy 
depends on the optical properties of the surface, in particular on the dispersion relation
of surface plasmon polaritons. 
The physical significance and the limits of validity of the obtained results are discussed in detail.
\end{abstract}


\maketitle

A striking feature of the quantum nature of the electromagnetic field is the existence of a vacuum energy and of zero-point fluctuations. Field fluctuations have observable effects such as the Lamb shift, spontaneous emission, Casimir-Polder forces, and the Casimir effect \cite{Milonni94,CPP95,Dalvit2011,Bordag2011}.
In particular, Casimir-Polder forces are long-range electromagnetic interactions between neutral polarizable atoms or molecules (van der Waals interaction), or between an atom and a macroscopic object, that arise from the zero-point fluctuations of the electromagnetic field and matter \cite{CP48,Craig1998,MPRSV08}.
Their prediction is arguably one of the most important results of quantum electrodynamics.
The macroscopic counterpart of the Casimir-Polder forces  is the Casimir effect, predicting an attractive force between two parallel uncharged plates \cite{Casimir1948}. The relation between the Casimir effect and Casimir-Polder forces has been extensively investigated in the literature, highlighting fundamental properties such as the nonadditivity \cite{Milonni94,Power1994} and fluctuations (see Ref. \cite{Messina2007} and references therein).
Experiments conducted from the second half of the nineties to now, have incontrovertibly shown the reality of these rather counterintuitive effects \cite{Sandoghdar1992,Sukenik1993,Dalvit2011,Bordag2011}.
It is now recognized that the Casimir effect plays an important role in the operation of micro- and nano-electromechanical devices (MEMS and NEMS) \cite{Serry1998,Dalvit2011,Bordag2011}. In particular, the research of experimental setups that allow for a modulation of the intensity or the sign of the force is of great interest as it may help to prevent the jamming (`stiction') of such machinery due to attractive forces, leading to the breakdown of the device. In the field of miniaturized atom traps, the control over the Casimir-Polder attraction is
desirable to achieve stable trapping at distances below one micron\cite{Salem2010}.

Recently, great attention has been paid to the so-called dynamic Casimir effect. Here, theory predicts that 
forcing  a plate to a rapid mechanical oscillatory motion leads to the generation of real photons from the electromagnetic vacuum state (see Ref. \cite{Dodonov2010} and references therein). The phenomenon has not yet been confirmed by experiments directly.
In fact, for a measurable number of real photons to be produced, the wall must  oscillate at frequencies that are not experimentally accessible, yet.
However, alternative schemes for the realization of the dynamic Casimir effect have been proposed. For example, the mechanical movement of the wall may be replaced by a suitable modulation of the optical properties of one of the surfaces \cite{Braggio07,Dodonov2006,Dodonov2006b} 
or of the optical path length of a cavity \cite{Dezael2010}.
Analogous effects have been recently observed in the context of superconducting circuits \cite{Johansson2009} and in trapped Bose-Einstein condensates \cite{Jaskula12}.

The microscopic counterpart of the dynamic Casimir effect is the dynamic Casimir-Polder effect
\cite{Westlund05, Shresta2003, Vasile2008, Messina2010}.
Recent work has studied the dynamic Casimir-Polder forces between an atom and a perfectly reflecting plate  \cite{Shresta2003, Vasile2008, Messina2010}.
It has been shown that after a sudden change of the atom-field interaction parameters,
the force exhibits oscillations in time and can be attractive or repulsive 
depending on time and the atom-wall distance. 
This is at variance with the stationary atom-plate Casimir-Polder force for a ground-state atom, 
which is generally attractive at all distances between the atom and the plate.
A description in terms of a perfectly reflecting plate 
obviously neglects the dynamics of charge transport \cite{Bordag2011, Bartolo2012} and 
does not take into account  
phenomena such as UV transparency and surface-mode excitations. In fact, these are known to play a crucial role 
in the near-field of conducting surfaces, and in particular in short-distance atom-surface interactions \cite{Annett1986, Ford1984, Failache2002, Henkel2002, Gorza2006, Stehle2011}.
We therefore generalize previous work on the dynamic Casimir-Polder interaction from the case of perfectly reflecting surfaces to allow for the description of an arbitrary surface material in terms of optical response functions.

More specifically, we investigate the dynamic (i.e., time-dependent) Casimir-Polder force between a neutral atom and a real surface 
and follow its relaxation towards a new equilibrium quantum state 
(``atom dressing'').
Such a situation can occur when a parameter involved in the atom-field coupling (such as the atomic transition frequency or the atom-plate distance) is suddenly modified \cite{Vasile2008, Messina2010}; in this case a time dependence of the force is expected.
We shall consider two different initial conditions for the system: an initially bare atom, and a partially dressed one \cite{Feinberg,PassanteVinci1996}.
The latter scenario occurs if a nonadiabatic change of the atomic properties is externally induced, for example by an external electric field.
We investigate the time-dependent atom dressing and the resulting atom-surface Casimir-Polder potential, and discuss how the physical properties of the surface material can influence the time dependence of the Casimir-Polder force, in particular its intensity and sign. This is expected, because the surface can respond to the field generated by the atom and 
participate in the dynamics of the system, in particular through surface excitations.
The results obtained in the two cases are compared with those already obtained in the literature for the limiting case of an ideal reflecting plate.
We find that also in the case of a general surface, the dynamic Casimir-Polder energy
exhibits oscillations in time, yielding a transient repulsive force, and 
that asymptotically settles to its stationary value.
Combining time-dependent perturbation theory with the matter-assisted field approach \cite{Knoll2001, Safari2006, Buhmann2007, Scheel2008, BuhmannBookI}, we evaluate the effects on the dynamic dressing 
caused by the metallic surface with a focus on the role of surface excitations. The formalism makes it possible to explore realistic situations close to those experimentally achievable, and it may also allow us to determine whether specific physical properties of the wall can be used to control the atomic dressing time and hence the intensity or the sign of the Casimir-Polder force.

The paper is organized as follows. In Sec. \ref{ch_Dyn-sec:bare_atom_dressing}, we give a brief review of the formalism used to calculate the time evolution of the atom-surface potential. The influence of material properties and surface mode excitations  on the dynamic atom-surface interaction is considered  in Sec. \ref{ch_Dyn-sec:material}. 
In Sec. \ref{sec:partialdressing}, we investigate the effect of a sudden change of the atomic properties on the dynamic atom-wall Casimir-Polder potential. 
Finally, in Sec. \ref{sec:Summary and Discussion}, we summarize and discuss the physical features and limits of validity of our results. Technical details related to the electromagnetic surface response (Green tensor)
are given in the Appendix.

%%%%%%%%%%%%%%%%%%%%%%%%%%%%%%%%%%%%%%%%%%
\section{Dynamic Dressing of an Initially Bare Atom}
\label{ch_Dyn-sec:bare_atom_dressing}
%%%%%%%%%%%%%%%%%%%%%%%%%%%%%%%%%%%%%%%%%%

%\label{ch_Dyn:spin_algebra}
Let us consider a neutral atom interacting with the quantum electromagnetic field in the presence of a real surface of arbitrary material.
The Hamiltonian of a polarizable atom interacting with the matter-assisted electromagnetic field in the multipolar coupling scheme and dipole approximation is \cite{Knoll2001, Safari2006, Buhmann2007, Scheel2008, BuhmannBookI}
\begin{eqnarray}
H 	&=& H_F + H_A + H_{AF}
\\
	&=&
	\int_0^\infty d\omega \int\!d^3 x\, 
	\hbar \omega \vec{f}^\dagger( \vec{x}, \omega) \cdot \vec{f}( \vec{x}, \omega)
\nonumber
\\
&& \quad {}
	+ \hbar \Omega S_z
	-\hat{\vec{d}}\cdot\vec{E}( \vec{r} )~,
\label{eq:hamiltonian}
\end{eqnarray}
where, for simplicity, the atom is described as a two level system, located at $\vec{r}$ and 
characterized by a single Bohr frequency $\Omega$ and pseudospin-operators $S_\pm, S_z$ \cite{CPP95}.
The operator
\begin{equation}
\hat{\vec{d}}
	=\vec{d} (S_+ + S_-)~,
\end{equation}
is the dipole moment, 
the vector $\vec{d}$ contains its transition matrix elements, 
assumed real, and takes the role of a coupling constant.
Moreover, $\vec{f}, \vec{f}^\dagger$ are generalized bosonic matter-field operators
with commutation relations
$[ \vec{f}( \vec{x}, \omega ),\, \vec{f}( \vec{x}', \omega' )] = 
\delta( \vec{x} - \vec{x}' ) \delta( \omega - \omega' )$.
The electric field operator $\vec{E}$ is expressed through the electric Green's tensor $\mathcal{G}_{mn}(\vec{x},\vec{x}', \omega)$ (see Appendix \ref{ch_App_em} for the explicit form) and reads \cite{Buhmann2007},
\begin{eqnarray}
E_m(\vec{r},t) &=&
	%	\sum_{\lambda}
	 \int_0^\infty d\omega \int d^3\vec{x}  \sqrt{\frac{\hbar }{\pi} \varepsilon_0 \Im \varepsilon(\vec{x},\omega)} \nonumber \\
	 && \times \left[\imath \mathcal{G}_{mn}(\vec{r},\vec{x}, \omega) f_n(\vec{x}, t; \omega) + h.c.\right]\,.
\label{ch_Dyn-eq:efield}
\end{eqnarray}

Throughout this work, Einstein sums over double spatial indices are implied.
The presence and the physical properties of the surface
are encoded in the Green's tensor
via reflection amplitudes and the dielectric function of the medium, $\varepsilon(\vec{x}, \omega)$.
When we make calculations for a specific surface model, we shall 
consider the simplest case of a nonmagnetic material with finite conductivity, that is the Drude model with a dielectric function
\cite{note}
\begin{eqnarray}
\varepsilon(\omega) = 1 - \frac{\omega_{\rm p}^2}{\omega(\omega + \imath \gamma)}~, \quad \mu(\omega)=1
\label{ch_Dyn-eq:Drude_model}
\end{eqnarray}
with $\omega_{\rm p}$ the plasma frequency and $\gamma$ the dissipation rate.

In the vicinity of a body of arbitrary dispersive and dissipative material, the excitations of the electromagnetic field cannot be separated from the excitations of the matter. This nonseparability is taken into account in the matter-assisted field approach \cite{Safari2006, Buhmann2007, Scheel2008} by the generalized bosonic operators $\vec{f}, \vec{f}^\dagger$.
Eq. \eqref{ch_Dyn-eq:efield} therefore generalizes a mode decomposition, necessary for the time-dependent perturbation theory which we will now employ.

Following the  procedure already used in previous works  \cite{Vasile2008, Messina2010}, the equation of motion for a general atomic or field operator $A$ is calculated from the Heisenberg equations. This leads to a series expansion $A(t) = \sum_{n} A^{(n)}(t)$, where $n$ indicates the power of the coupling constant. Up to the first order in the coupling, we find
\begin{eqnarray}
	\label{ch_Dyn-eq:atom_op_order0}
S_+^{(0)}(t)
	&=&  S^{(0)}_+(0)\EXP{\imath \Omega t}~,\\
S_+^{(1)}(t)
	&=& \frac{2 \imath }{\hbar} \EXP{\imath \Omega t} \int_0^t dt' S_z^{(0)}(0) \vec{d}\cdot \vec{E}^{(0)}(t') \EXP{-\imath \Omega t'}~,
\end{eqnarray}
and
\begin{eqnarray}
\vec{f}^{\dagger(0)}(\vec{x},t; \omega)
	&=&\vec{f}^{\dagger}(\vec{x},0; \omega) \EXP{\imath\omega t}~,\\
f_n^{\dagger(1)}(\vec{x},t;\omega)
	&=& \EXP{\imath \omega t}
	\sqrt{\frac{\varepsilon_0 }{\hbar \pi} \Im \varepsilon(\omega)}
	 d_m \mathcal{G}_{mn}(\vec{x}, \vec{r}, \omega)
	  \nonumber\\
	 &&
	\hspace{-2cm}\times
	 \left[S^{(0)}_+(0) F(\Omega - \omega,t) +S^{(0)}_-(0)F(-\Omega -\omega, t) \right]\,.~%\nonumber\\
	\label{ch_Dyn-eq:field_op_order1}
	%\\
\end{eqnarray}

We stress that we do not impose the rotating wave approximation (RWA) so that Eq. \eqref{ch_Dyn-eq:field_op_order1}
 contains both resonant (first term) and antiresonant (second term) contributions. The time-dependence is abbreviated by
\begin{equation}
F(\omega, t) = \int_0^t dt' \EXP{\imath \omega t'} = \frac{\EXP{\imath \omega t} - 1}{\imath \omega}~.
\end{equation}
The solutions obtained above are operator equations, valid for any initial state.

We now consider a specific initial state at $t=0$ 
for the system. 
In particular, we consider a ground-state atom at zero temperature.
The time-dependent Casimir-Polder potential is obtained from the energy shift
$ U(z, t) 	= \frac{1}{2} \langle H_{AF} (t)\rangle$ and, by symmetry, can depend only on the distance $z$ between the atom and the surface.
 Our analysis is restricted to the second order in the coupling constant $\vec{d}$ and mean values are taken with respect to the bare ground-state $| \mathrm{vac}, \downarrow\rangle$, so that
\begin{align}
\label{ch_Dyn-eq:energy_shift_groundstate}
U(z, t) \approx&-  \textstyle \frac{1}{2} \langle \hat{\vec{d}}^{(0)} \cdot \vec{E}^{(0)} \rangle \nonumber
\\&		- \textstyle \frac{1}{2} \langle \hat{\vec{d}}^{(0)} \cdot \vec{E}^{(1)} \rangle
- \textstyle \frac{1}{2} \langle \hat{\vec{d}}^{(1)} \cdot \vec{E}^{(0)} \rangle~.
\end{align}

It is clear that the first term is off-diagonal in the spin basis and does not contribute. The remaining terms can be evaluated by using the identity
\begin{align}
& \Im{\mathcal G}_{ij}(\vec r_1, \vec r_2, \omega) = \nonumber
\\
& \int d^3 \vec{x} \varepsilon_0 \Im[\varepsilon(\omega, \vec{x})]{\mathcal G}_{ik}(\vec{r}_1, \vec{x}, \omega){\mathcal G}^*_{jk}(\vec{x},\vec{r}_2, \omega)~,
\label{ch_Dyn-eq:knoells_magical_formula}
\end{align}
which is a known property of the electromagnetic Green's tensors contained already in Maxwell's equations \cite{Eckhardt1984, Dung1998, Knoll2001}. Besides, from the reciprocity principle we have $\mathcal{G}_{ij}(\vec{r},\vec{r}',\omega) = \mathcal{G}_{ji}(\vec{r}',\vec{r},\omega)$ and for hermitean fields $\mathcal{G}_{ij}(\vec{r},\vec{r}',-\omega)=\mathcal{G}^*_{ij}(\vec{r},\vec{r}',\omega)$ (with $\omega \in \mathbbm{R}$).
Thus,
 \begin{widetext}
\begin{align}
	\label{ch_Dyn-eq:energy_shift_term2}
 - \frac{1}{2} \langle \hat{d}^{(0)}_m E^{(1)}_m (\vec r,t) \rangle &=
- \imath d_m d_n \int_0^\infty \frac{d\omega}{2 \pi} \Im \mathcal G_{mn}(\vec r,\vec r,\omega)
%\times \nonumber\\
%	&\times
	\left[\EXP{-\imath (\Omega+\omega)t } F(\Omega + \omega, t) - \EXP{-\imath (\Omega- \omega)t} F(\Omega - \omega,t)\right]
%\nonumber
\\
- \frac{1}{2} \langle \hat{d}^{(1)}_m  E^{(0)}_m (\vec r,t) \rangle &=
	\label{ch_Dyn-eq:energy_shift_term3}
 + \imath d_m d_n   \int_0^\infty \frac{d\omega}{2 \pi}	\Im \mathcal G_{mn}(\vec r,\vec r,\omega)
 %\times \nonumber\\	& \times
 \left[ \EXP{\imath (\Omega + \omega)t} F(-\Omega - \omega, t) - \EXP{-\imath(\Omega-\omega)t} F(\Omega - \omega,t)\right]~.
%	\nonumber
	\end{align}
\end{widetext}
Summing up Eqs. \eqref{ch_Dyn-eq:energy_shift_term2} and 
\eqref{ch_Dyn-eq:energy_shift_term3}, 
the last terms with a pole at $\omega = \Omega$ 
cancel out each other and the energy shift at second order becomes
\begin{eqnarray}
U(\vec{r},t) %&=&
		&=&  -d_m d_n   \int_0^\infty \frac{d\omega}{2 \pi}	
		\Im \mathcal{G}_{mn}(\vec r,\vec r,\omega) \nonumber \\
		&&
		\times  \left ( \frac 2{\Omega + \omega}- \frac {2 \cos\left[ (\Omega + \omega)t \right]}{\Omega + \omega} \right)
	\label{ch_Dyn-eq:energy_shift_complete}.
\end{eqnarray}
From now on, we will use the shorthand
\begin{equation}
G(\omega) \equiv d_m d_n \mathcal{G}_{mn}( \vec{r}, \vec{r}, \omega)
\label{eq:def-shorthand-G}
\end{equation}
for the projected single-point Green's tensor.
The previous Eq.~\eqref{ch_Dyn-eq:energy_shift_complete}
is easily recognized as the equivalent of Eq.~(13) of Ref. \cite{Vasile2008} and it is manifestly real.
In order to distinguish the static contribution from the time-dependent term, we separate the two terms in the second line of Eq. \eqref{ch_Dyn-eq:energy_shift_complete} and define
\begin{equation}
U(\vec{r}, t) = U_{\rm stat}(\vec{r}) + U_{\rm dyn}(\vec{r}, t)~.
\label{ch_Dyn-eq:energy_shift_contributions_def}
\end{equation}

The system is unperturbed at $t=0$ and its evolution must be continuous so that $U(\vec{r}, 0) = 0$. The correction $U_{\rm dyn}$ describes the transient dressing dynamics and averages to zero in the stationary limit \mbox{$t \to \infty$}, due to the oscillatory integrand. In this limit, we recover the stationary Casimir-Polder potential $U_{\rm stat}$ given in Ref. \cite{Wylie1985}.
Note that expression \eqref{ch_Dyn-eq:energy_shift_complete} does not depend any more explicitly on a bulk property like $\Im \varepsilon(\vec{x},\omega)$,
but only on the total Green's tensor above the surface.
This allows for the calculation of the time-dependent energy shift of an atom in a rather general setting, and remains valid even in cases where the microscopic details of the surface response are completely ignored, as in the limit case of a perfect reflector.
In the following we shall use expression \eqref{ch_Dyn-eq:energy_shift_complete}  to obtain the dynamic atom-wall Casimir-Polder potential,  for an atom placed near a planar surface described by a dielectric or metallic half-space. 

As a first step, we now consider the case of ideal reflector;
the dynamic atom-wall Casimir-Polder energy has been already investigated in Refs. \cite{Vasile2008} and provides a convenient cross-check. 
The two-level atom is assumed to be isotropically polarizable, 
so that $d_m d_n = \frac{1}{3}|d|^2 \delta_{mn}$.
In the limit of a perfect reflector (reflection coefficients $r_{\rm p} = -r_{\rm s} = 1$), the single-point Green's tensor may be expressed through a spatial differential operator $\mathcal{D}$:
\begin{eqnarray}
\label{ch_Dyn-eq:GT_perf}
% d_m d_n
% %\omega^2
% \mathcal G_{mn}(\omega) 
 G( \omega )
 &\mapsto& \frac{|d|^2}{3} \mathcal{D} \, \EXP{2 \imath \omega z \alpha / c}~,\\
 \mathcal{D} &=& \lim_{\alpha \to1}\frac{1}{16 \pi \varepsilon_0 z^3} \left( 2 - 2 \frac{\partial }{\partial \alpha} + \frac{\partial^2 }{\partial \alpha^2}\right).
\end{eqnarray}
We immediately obtain
\begin{align}\nonumber
U(z,t) =& -\frac{|d|^2}{3}\int_0^\infty \frac{d\omega}{2 \pi}
\mathcal{D} \sin(2 \alpha z  \omega/ c )\\
&\times
		 \frac{2 - 2 \cos\left[ (\Omega + \omega)t \right]}{\Omega + \omega}
	\label{ch_Dyn-eq:energy_shift_perfect}~,
\end{align}
where $z$ is the atom's distance with respect to the plate.
This equation corresponds to Eq.\,(13) of Ref. \cite{Vasile2008}. There, an explicit expression for the force was obtained in terms of cosine and sine integral functions.
%================================================
\begin{figure}[t!]
\centering
 \includegraphics[height=5cm]{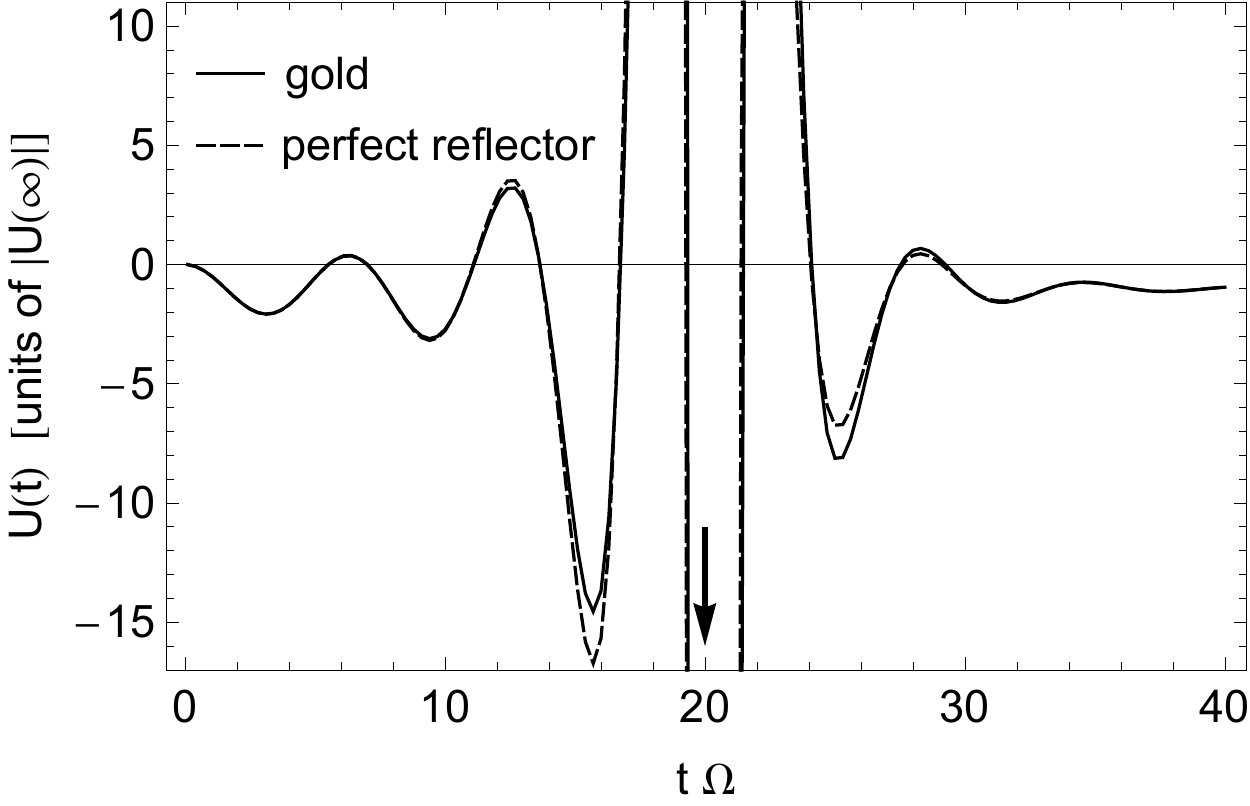}
 \hspace{1ex}
\caption{Time dependent energy shift near a gold surface (solid line) and near a perfect reflector (dashed line).
Parameters for gold are
$\hbar\omega_{\rm p} = 8.9\,\unit{eV}$, % = 1.35 \times 10^{16} \unit{rad~s}^{-1}$,
$1/\gamma = 19\,\unit{fs}$. %$\gamma = 4 \times 10^{-3} \omega_{\rm p}$.
The atomic transition frequency 
corresponds to the $780\,\unit{nm}$ line of rubidium 85
($\Omega % = 2.4\times 10^{15} \unit{rad~s^{-1}} 
= 0.18\,\omega_{\rm p}$),
and the atom-surface distance is $z = 10 \,c/\Omega = 1.2\,\unit{\mu m}$.
At the signal-transit time $\Omega t = 20$, indicated by the arrow, the reflected light cone reaches the atom. On this scale, the curve obtained from the Drude model for gold and from its lossless limit ($\gamma \to 0$), respectively, cannot be distinguished.
}
\label{ch_Dyn-fig:dynamic_CP}
\end{figure}
%================================================
A numerical evaluation of Eq. \eqref{ch_Dyn-eq:energy_shift_perfect}, giving the time-dependent energy shift of an atom near a perfect reflector, is shown in Fig. \ref{ch_Dyn-fig:dynamic_CP} (dashed line).
Important features of Fig. \ref{ch_Dyn-fig:dynamic_CP} have already been discussed in Refs. \cite{Shresta2003, Vasile2008}. It is shown that the dynamic atom-wall Casimir-Polder energy, exhibits oscillations in time, yielding transient repulsive force and that in the asymptotic limit it settles to a stationary value. Strong effects set in when the atom begins to `see' its own mirror image, precisely at the signal-transit time $t = 2 z/c$, when the reflected light cone reaches the atom. 
As explained in Refs. \cite{Shresta2003, Vasile2008}, 
the nontrivial divergence in the radiation reaction at this moment can be traced back to the point dipole approximation \cite{Milonni1976} and to assuming the atom initially bare \cite{PassanteVinci1996}.
The energy shift shows a precursor for $t < 2 z / c$
because the atom starts to evolve unitarily under the full Hamiltonian, i.e. it couples instantaneously to the fluctuating 
field~\cite{Vasile2008}.

In the next Section we shall consider in detail the dressing problem and the atom-wall Casimir-Polder interaction in the case of a general surface, and we discuss in detail the role of the physical properties of the surface on the dynamic atomic dressing.

%%%%%%%%%%%%%%%%%%%%%%%%%%%%%%%%%%%%%%%%%%%
\section{Material Properties and Surface Mode Excitations}
\label{ch_Dyn-sec:material}
%%%%%%%%%%%%%%%%%%%%%%%%%%%%%%%%%%%%%%%%%%
Let us now consider an atom near a surface with an arbitrary linear electromagnetic response.
Here, the evaluation of Eq. \eqref{ch_Dyn-eq:energy_shift_complete} is not straightforward because of the oscillating integrand. It is therefore very useful to perform an analytic continuation to complex frequencies as shown in the following subsection.
We take advantage of the fact that the Green function $G( \omega )$ is 
analytic for complex $\omega$ in the upper half-plane, as it must for a retarded
response.

%%%%%%%%%%%%%%%%%%%%%%%%%%%%%%%%%%%%%%%%%%
\subsection{Imaginary frequency representation}
\label{ch_Dyn-sec:imaginary_frequencies}
%%%%%%%%%%%%%%%%%%%%%%%%%%%%%%%%%%%%%%%%%%

\begin{figure}[t!]
\hspace{-2cm}  \includegraphics[height=5cm]{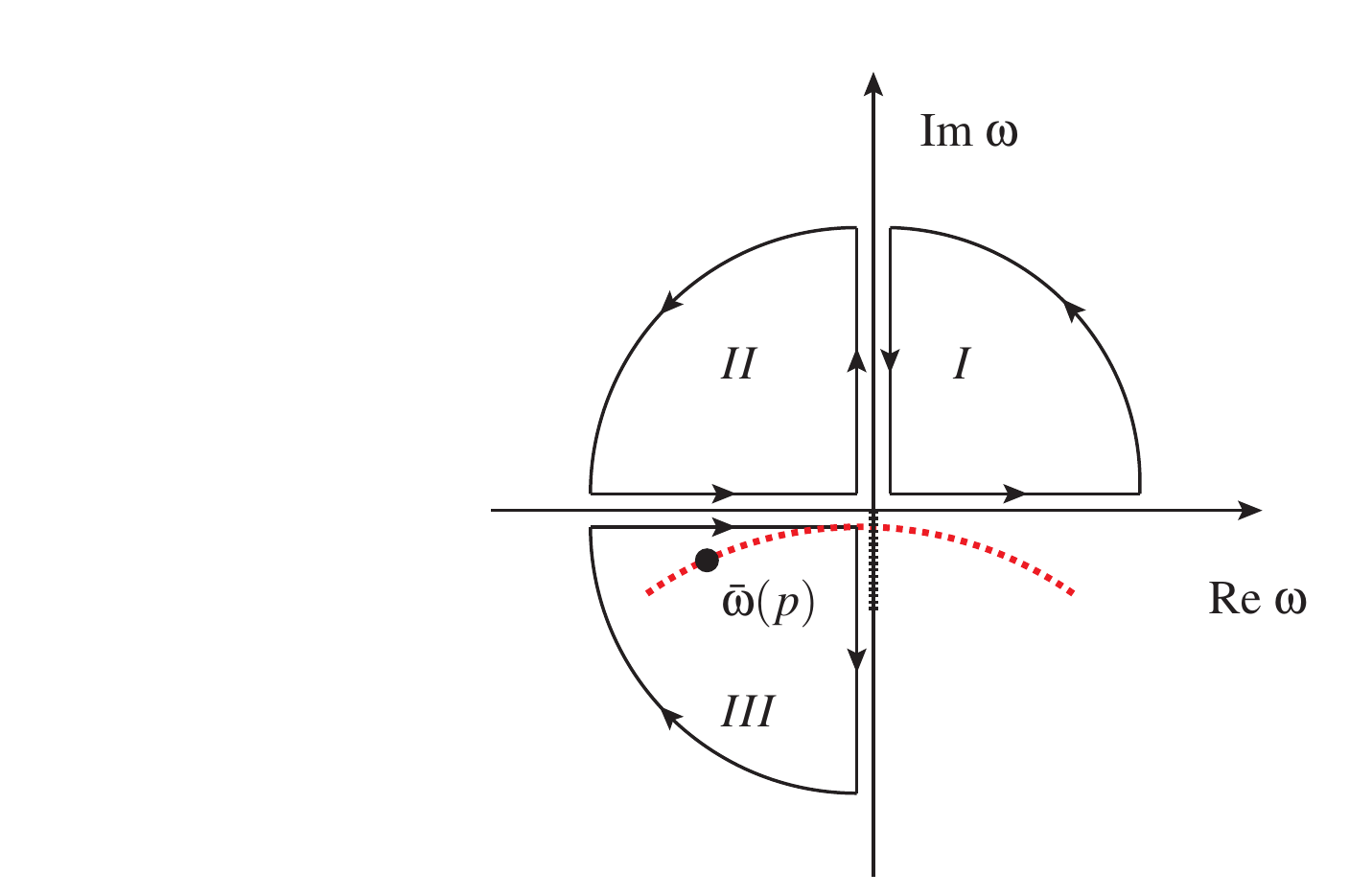}%\\
\caption{(Color online) Integration contours for the analytic continuation. In the range $\Re \omega > 0$, convergence is always guaranteed for the first integral in \eqref{ch_Dyn-eq:energy_shift_stat_im} in the upper half-space (contour I). For the second integral in \eqref{ch_Dyn-eq:energy_shift_stat_im}: outside the reflected light cone ($c t < 2 z$), the integration contour  $\Re \omega <0 $ is closed in the upper complex plane (contour II); inside the light cone ($c t > 2 z$), contour III applies, avoiding a possible branch-cut along negative imaginary axis (hatched line).
The red dotted line symbolizes the surface-plasmon resonances at $\bar{\omega}(p)$, contributing a single isolated pole at a given value 
of momentum $p$ parallel to the surface.
}
\label{ch_Dyn-fig:contour}
\end{figure}

The static Casimir-Polder potential is often written as an integral over 
imaginary frequencies. 
To achieve this form, 
we use the relation $2 \imath \, \Im G(\omega) =  G(\omega) - G(- \omega) $ ($\omega \in \mathbbm{R}$) and integrate along contours I and II in Fig. \ref{ch_Dyn-fig:contour}. 
Using the fact that the surface becomes transparent as 
$|\omega| \to \infty$ in the upper half-plane,
we recover the expression for the static shift reported in Ref. \cite{Wylie1985},
\begin{align}
\label{eq:CP_stat}
U_{\rm stat}(z) =
		 -  % d_m d_n   
		 \int_0^\infty \frac{d\xi}{2 \pi}
		 %\xi^2
%		  \mathcal{G}_{mn}(\imath \xi)
		  G(\imath \xi)
		\frac{2 \Omega}{\Omega^2 + \xi^2}.
\end{align}
The analytic properties of the dynamic part are more subtle. Expressing the cosine in Eq. \eqref{ch_Dyn-eq:energy_shift_complete}
as the real part 
of a complex exponential, we find
\begin{eqnarray}
\label{ch_Dyn-eq:energy_shift_stat_im0}
U_{\rm dyn}(z,t)
%		&& = % d_m d_n  
%		\Re \frac{2  \EXP{\imath \Omega t}}{2 \imath}
%		   \int_0^\infty \frac{d\omega}{2 \pi} \frac{\EXP{\imath \omega t}}{\omega + \Omega}
%		  \left[
%		%   \mathcal{G}_{mn}( \omega ) - \mathcal{G}_{mn}( -\omega)
%		G( \omega ) - G( -\omega)
%		\right]
%		\nonumber\\
% \label{ch_Dyn-eq:energy_shift_stat_im}
		=&&- % d_m d_n  
		\Re \imath \,\EXP{\imath \Omega t}
		   \int_0^\infty \frac{d\omega}{2 \pi}
		 %  \mathcal{G}_{mn}(\omega) 
		   G(\omega) 
		   \frac{\EXP{\imath \omega t}}{\omega + \Omega}
		  \nonumber\\
		    &&+
		   % d_m d_n  
		   \Re \imath \,\EXP{\imath \Omega t} \int_{-\infty}^0 \frac{d\omega}{2 \pi}
		%   \mathcal{G}_{mn}(  \omega)
		   G(  \omega)
		\frac{\EXP{-\imath \omega t}}{-\omega + \Omega}.
\label{ch_Dyn-eq:energy_shift_stat_im}
\end{eqnarray}
In the first line, the integrand is regular in the upper complex 
half-plane and can be shifted to the positive imaginary axis (contour I 
in Fig. \ref{ch_Dyn-fig:contour}).
Two different cases must be considered for the second integral, according to times before or after the  transit time $2z/c$.
As can be easily inferred from the expressions obtained in the Appendix,
the asymptotic behavior of the Green's tensor for large values of $|\omega|$ is of the form $G(z, \omega) \EXP{- \imath \omega t} = \EXP{\imath \omega (2 z/ c - t)} g(z, \omega)$.
If $ct<2z$, i.e. before the signal transit time,  the integral converges in the upper complex half-plane;  if $c t > 2z$, i.e. afterwards, it converges in the lower complex half-plane. The integrals can therefore be evaluated using the contours II and III in Fig. \ref{ch_Dyn-fig:contour}, respectively.
Note that for ground-state atoms, the poles in $\Omega$ are never inside 
the contour, so that there are no contributions from atomic resonances.
The time-dependent potential is
\begin{eqnarray}
U_{\rm dyn}(z,t)%\nonumber\\
=&& % d_m d_n  
	\Re \EXP{\imath \Omega t}  %\nonumber\\
  \biggl[
				 	\int_0^\infty \frac{d \xi}{2 \pi}
					%\xi^2
		% \mathcal G_{mn}(\imath \xi) 
		G(\imath \xi) 
		\frac{\EXP{- \xi t}}{\imath \xi + \Omega} \nonumber\\
	&&+	\int_0^\infty \frac{d \xi}{2 \pi}
	%\xi^2
	% \mathcal G_{mn}(\imath \xi) 
	G(\imath \xi) 
	\frac{\EXP{ \xi t}}{-\imath \xi + \Omega}
	 \theta(2z - c t)\nonumber\\
	&&- 	\int_{-\infty}^0 \frac{d \xi}{2 \pi}
	%\xi^2
	% \mathcal G_{mn}(\imath \xi) 
	G(\imath \xi) 
	\frac{\EXP{ \xi t}}{-\imath \xi + \Omega}
	\theta(c t-2z)\nonumber\\
	&& +  \imath \oint_{III} \frac{d \omega}{2 \pi}
	%\omega^2
	%\mathcal G_{mn}(\omega) 
	G( \omega ) 
	\frac{\EXP{- \imath \omega t}}{- \omega + \Omega} \theta(c t-2z)
		 \biggr].\hspace{3ex}
		 \label{ch_Dyn-eq:energy_shift_dyn_im}
\end{eqnarray}
This expression is a central result of the present work.
The imaginary frequency integrals are sufficiently well-behaved to be evaluated numerically using a general Green's tensor.
The last term in Eq. \eqref{ch_Dyn-eq:energy_shift_dyn_im} will be denoted by $\Delta U_{\rm res}$ and includes the transient contribution due to material excitations (poles of the Green's tensor), 
as we shall discuss 
in detail in the next subsection.

%================================================
%================================================

To get rapidly converging expressions,  we have so far written all integrals over retarded Green's tensors (positive frequency arguments) which are analytic in the upper complex plane.
However,  from the form of Eq. \eqref{ch_Dyn-eq:energy_shift_stat_im0} only odd-parity terms of the Green's tensor contribute. In fact we may obtain equivalent results by considering $\left[G(\omega) - G(-\omega)\right] e^{\imath \omega t}$ a meromorphic function that converges along the contour I for $t> 2z/c$ but contains a pole in the upper half-plane.

A numerical evaluation of the total time-dependent energy shift obtained from the imaginary frequency formulation of Eqs. \eqref{ch_Dyn-eq:energy_shift_stat_im} and \eqref{ch_Dyn-eq:energy_shift_dyn_im} is compared in Fig. \ref{ch_Dyn-fig:dynamic_CP}  to the results for a perfect reflector.
Clearly, for this distance in the far zone ($z \gg c / \Omega$), 
the dependence on the material properties is relatively weak.
This is because in this regime, 
the electromagnetic field, except very close to the reflected light cone, is mediated by propagating photonic modes with low frequency: for these modes the perfect reflector provides a reasonable boundary condition even for a real material. In fact, this can be shown directly for the Green's tensor 
at large distances above a general metal \cite{Haakh2009b}.
As shown in Fig. \ref{ch_Dyn-fig:dynamic_CP},
the potentials obtained for a perfect reflector and for a gold surface differ strongest at times close to the signal transit time: 
at this instant, the response is dominated by
high-energy modes, and a gold surface shows a cutoff above its 
plasma frequency.

%%%%%%%%%%%%%%%%%%%%%%%%%%%%%%%%%%%%%%%%%%
\subsection{Surface-plasmon contribution}
\label{sec:material_resonances}
%%%%%%%%%%%%%%%%%%%%%%%%%%%%%%%%%%%%%%%%%%

%================================================
\begin{figure}[b!]
\centering  \includegraphics[height=5cm]{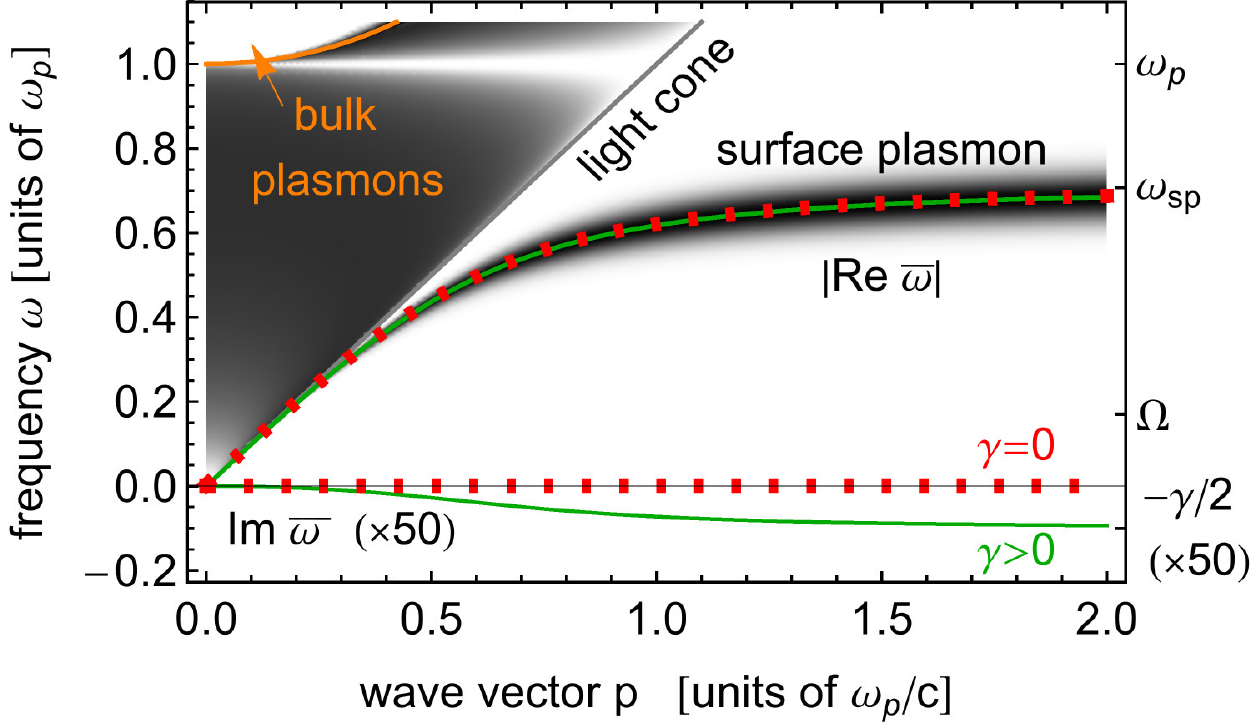}%\\
\caption{
(Color online) Surface plasmon dispersion relation $\bar{\omega}( p )$
in the Drude model for gold  (green solid curves,
parameters as in Fig. \ref{ch_Dyn-fig:dynamic_CP}).
Red dotted curves correspond to the lossless limit  $\gamma \to 0$ of Eq. \eqref{ch_Dyn-eq:SPDR}. The plasmonic frequency scales
$\omega_{\rm p}$ and $\omega_{\rm sp}$, and the atomic 
resonance frequency $\Omega$ are indicated for the parameters
of Fig.\ref{ch_Dyn-fig:dynamic_CP}. 
The gray scale background gives $|\Im r_{\rm p}(\omega, p)|$.
}
\label{ch_Dyn-fig:spdr}
\end{figure}
%================================================
The matter-assisted photon modes (plasmons or polaritons) near 
a real material are encoded in
the analytic structure of the Green function $G( \omega )$. In particular
surface excitations play an important role \cite{Flores1979, Raether1988, %
Ford1984}. They appear as poles and singularities of $G( \omega )$ in the 
lower half-plane $\Im \omega < 0$. For times $t > 2 z / c$, they contribute
to the time-dependent potential, as can be seen from the last integral 
in Eq.\eqref{ch_Dyn-eq:energy_shift_dyn_im}.
Different types of polaritons can be identified from the integral
representation of $G( \omega )$ given in Eq.(\ref{ch_App_em-eq:mag_greentensor})
of the Appendix.
Isolated poles of the reflection coefficients correspond to surface plasmon modes. 
The dielectric function $\varepsilon( \omega )$ gives rise to bulk plasmons 
in the high-frequency range $\omega > \omega_p$. 
A conducting medium features additional diffusive low-frequency excitations (`eddy currents') connected with branch-cut discontinuities along the negative imaginary axis.
Since the latter have a minor impact 
in electric-dipole coupling \cite{Haakh2009b, Intravaia2009, Intravaia2010},
we focus in the following on the surface excitations.

A surface described by the Drude model of Eq. \eqref{ch_Dyn-eq:Drude_model} carries a single surface plasmon given by a pole of the reflectivity in p-polarization
\begin{align}
\varepsilon(\omega)\kappa(\omega, p) + \kappa_{m}(\omega, p) = 0
	~,
\end{align}
where the notation of Appendix~\ref{ch_App_em} is used.
The solution $\bar{\omega}(p)$ is shown in Fig. \ref{ch_Dyn-fig:spdr} and will be used for numerical evaluations in Sec. \ref{ch_Dyn-sec:near_field_dressing}.
In the general case, the analytic form is not very instructive. However, a simple expression is obtained in the lossless limit $\gamma \to 0$. Here, the relevant branch of the surface-plasmon dispersion relation 
 inside the contour III of Fig. \ref{ch_Dyn-fig:contour} is \cite{Raether1988}
\begin{align}
\label{ch_Dyn-eq:SPDR}
\bar{\omega}(p) &= - \sqrt{ \omega_{\rm sp}^2 +  c^2 p^2 - \sqrt{\omega_{\rm sp}^4 + c^4 p^4} }  - \imath 0^+~,\\
\label{ch_Dyn-eq:SPDR-kappa}
% \bar{\kappa}(p) &= \sqrt{p^2 - \bar{\omega}^2(p) / c^2}~,
\bar{\kappa}(p) &= \sqrt{ \sqrt{\omega_{\rm sp}^4/c^4 + p^4} - \omega_{\rm sp}^2/c^2 }~,
\end{align}
where the surface-plasmon frequency $\omega_{\rm sp} = \omega_{\rm p} / \sqrt{2}$ is the asymptotic value at large wave vectors
and $\bar{\kappa}(p)$ governs the decay of the surface mode outside the
material.
The surface plasmon-polariton is an evanescent wave, i.e. its dispersion relation lies below the light cone 
[$\bar{\omega}(p) < p c$, see Fig. \ref{ch_Dyn-fig:spdr}]. 
Fig. \ref{ch_Dyn-fig:spdr} illustrates that Eq. \eqref{ch_Dyn-eq:SPDR} provides
a very good approximation to the resonance frequency $\Re \bar\omega$
of a good conductor ($\gamma \ll \omega_{\rm p}$).
Of course, it cannot account adequately for plasmon damping ($\Im \bar\omega$). 
We will see that damping plays a role at short distances 
(Sec.\ref{ch_Dyn-sec:near_field_dressing}) and gives the otherwise 
monochromatic dispersion relation a finite width \cite{Joulain2005}.

\begin{figure}[t!]
{
\centering
 \includegraphics[height=5cm]{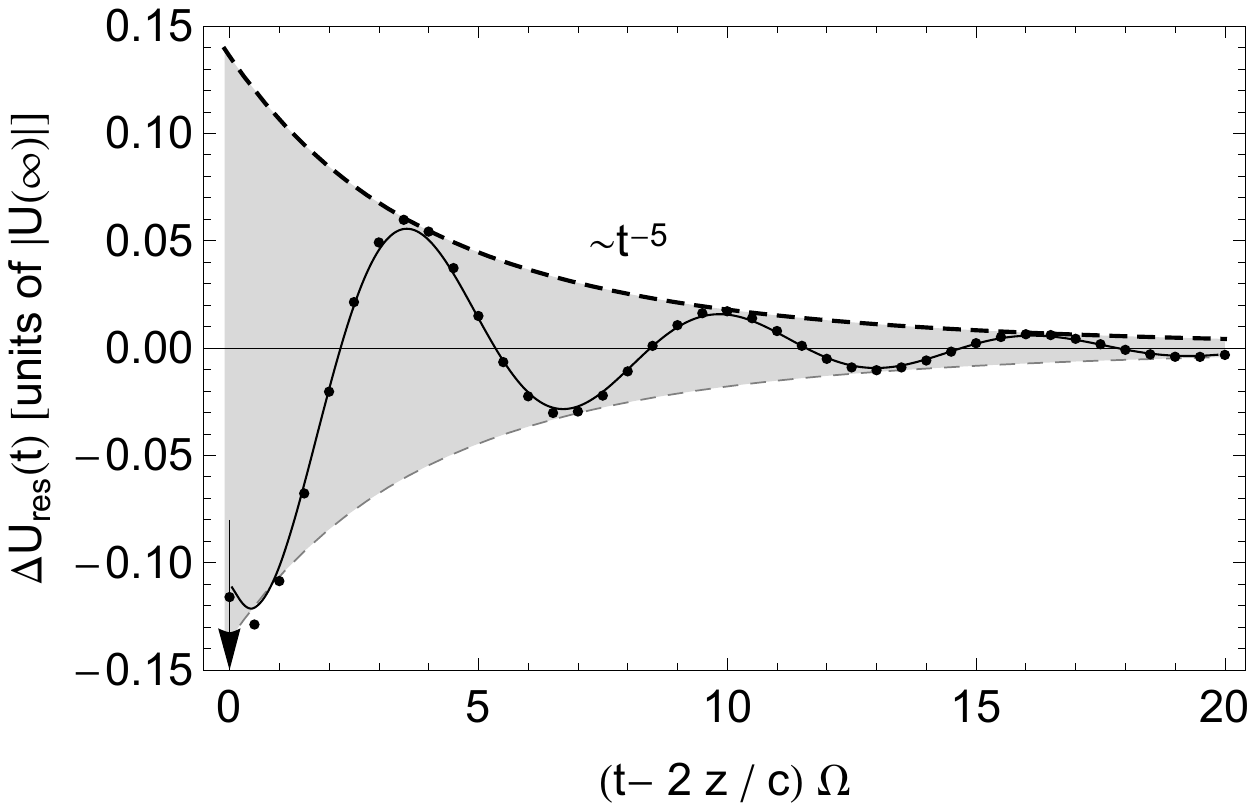}
\caption{
Resonant surface mode contribution $\Delta U_{\rm res}$ [Eq. \eqref{ch_Dyn-eq:pole_contribution4}] to the atom-surface dressing near a gold surface (solid line,  parameters as in Fig. \ref{ch_Dyn-fig:dynamic_CP}. Dots give the lossless limit $\gamma \to 0$). 
Relatively large atom-surface distance:  $z = 10 c/ \Omega$. 
The contribution sets in only after the signal-transit time $t = 2 z /c$. The envelope shows good agreement with the $t^{-5}$ power law expected at this distance}
\label{ch_Dyn-fig:decay_time}
}
\end{figure}

Using the integral form \eqref{ch_App_em-eq:mag_greentensor} of the electric Green's tensor  
and the previous expression for the plasmon dispersion, it is possible to calculate separately the resonant contribution in Eq. \eqref{ch_Dyn-eq:energy_shift_dyn_im}, similar to previous approaches in the static case \cite{Annett1986}.
The integrals over wave vectors and frequency are interchanged and the atom is assumed to be isotropically polarizable, so that ($t>2z/c$)
\begin{align}
\Delta & U_{\rm res}(z,t)	 \nonumber \\
		=& \frac{|d|^2 }{12 \pi \varepsilon_0}
		\Re \imath \,\EXP{\imath \Omega t}
		\int_0^\infty dp \oint_{III} \frac{d \omega}{2 \pi} \frac{\EXP{- \imath \omega t - 2 \kappa z}}{ -\omega + \Omega}  \frac{p}{\kappa} \nonumber\\
		&\times \left[\frac{\omega^2}{c^2} r_{\rm s}(\omega, p) + r_{\rm p}(\omega, p)\left(\frac{\omega^2}{c^2} + 2 \kappa^2\right)\right]~.%\nonumber
		\label{ch_Dyn-eq:pole_contribution3}
\end{align}
The surface plasmon arises from a single isolated pole at 
$\omega = \bar{\omega}(p)$ which 
is located inside the contour due to its negative imaginary part. In the lossless limit of Eq. \eqref{ch_Dyn-eq:SPDR}, this is achieved by the infinitesimal imaginary shift.
We calculate the residue from a series expansion of the denominator of the $r_{\rm p}$ coefficient, 
\begin{eqnarray}
\label{ch_Dyn-eq:Residue}
R_{\rm p}(\omega, p) =
\frac{\varepsilon(\omega) \kappa - \kappa_m}{\frac{d}{d\omega} [\varepsilon(\omega) \kappa + \kappa_m]}~.
\end{eqnarray}
Using the residue theorem for the clockwise contour, the contribution due to the surface plasmon, present at  $t > 2z/c$, is
\begin{eqnarray}
 \label{ch_Dyn-eq:pole_contribution4}
 \Delta U_{\rm res}(z, t)	&=& \frac{|d|^2}{12 \pi \varepsilon_0} \Re \frac{\EXP{\imath \Omega t}}{2}
 \int_0^\infty dp  \frac{p}{\bar{\kappa}(p)}
 \\
&& \hspace{-1.5cm}\times
\frac{\EXP{- \imath \bar{\omega}(p) t - 2 \bar{\kappa}(p) z}}{ -\bar{\omega}(p) + \Omega}   \left(2  p^2 - \frac{\bar{\omega}^2(p)}{c^2}\right)  R_{\rm p}(\bar{\omega}(p), p)\,.
				\nonumber
\end{eqnarray}
This resonant contribution to the atom-surface interaction is plotted in
Fig. \ref{ch_Dyn-fig:decay_time},
normalized to $U(\infty )$ near a gold surface. 
The full complex dispersion relation
is nearly indistinguishable from the lossless limit discussed above.

%================================================
%================================================

\subsection{Large distance}

The data of Fig. \ref{ch_Dyn-fig:decay_time} are calculated for a distance
larger than the atomic resonant wavelength (retarded regime). The surface
modes then contribute only a minor fraction of the total dynamic Casimir-Polder
energy.
Their contribution is a transient effect 
and decays quickly in time due to the wide frequency range in the
integral  \eqref{ch_Dyn-eq:energy_shift_contributions_def}.
In the far field, where small values of $p$ contribute and where the plasmon
dispersion relation approaches the light cone, even a dissipative medium provides very little damping (cf. Fig. \ref{ch_Dyn-fig:spdr}).
The relaxation is, therefore, clearly not a dissipative effect but rather 
due to the dephasing between spectral components, because
of the broad bandwidth of the plasmon dispersion covering the interval 
$|\bar{\omega}(p)|\in [0;\omega_{\rm sp}]$.

We get some insight by studying 
the envelope of the resonant potential in the retarded regime.
The relevant contributions to the integral in Eq. \eqref{ch_Dyn-eq:pole_contribution4} stem 
from low-momentum modes near the light cone. 
We change from $p$ to $\bar\omega$ as integration variable using the 
dispersion relation [Eqs. \eqref{ch_Dyn-eq:SPDR}, \eqref{ch_Dyn-eq:SPDR-kappa}] 
and expand the integrand for small frequencies. This gives a power law $\bar{\omega}^4$
and therefore an algebraic decay $\propto t^{-5}$ at large times.
The good agreement between this power law and the numerical solution is shown 
in Fig. \ref{ch_Dyn-fig:decay_time}.

%%%%%%%%%%%%%%%%%%%%%%%%%%%%%%%%%%%%%%%%%%
\subsection{Intermediate distance}
\label{sec:intermediate-dist}
%%%%%%%%%%%%%%%%%%%%%%%%%%%%%%%%%%%%%%%%%%
%================================================
\begin{figure}[t!]
\includegraphics[height=5cm]{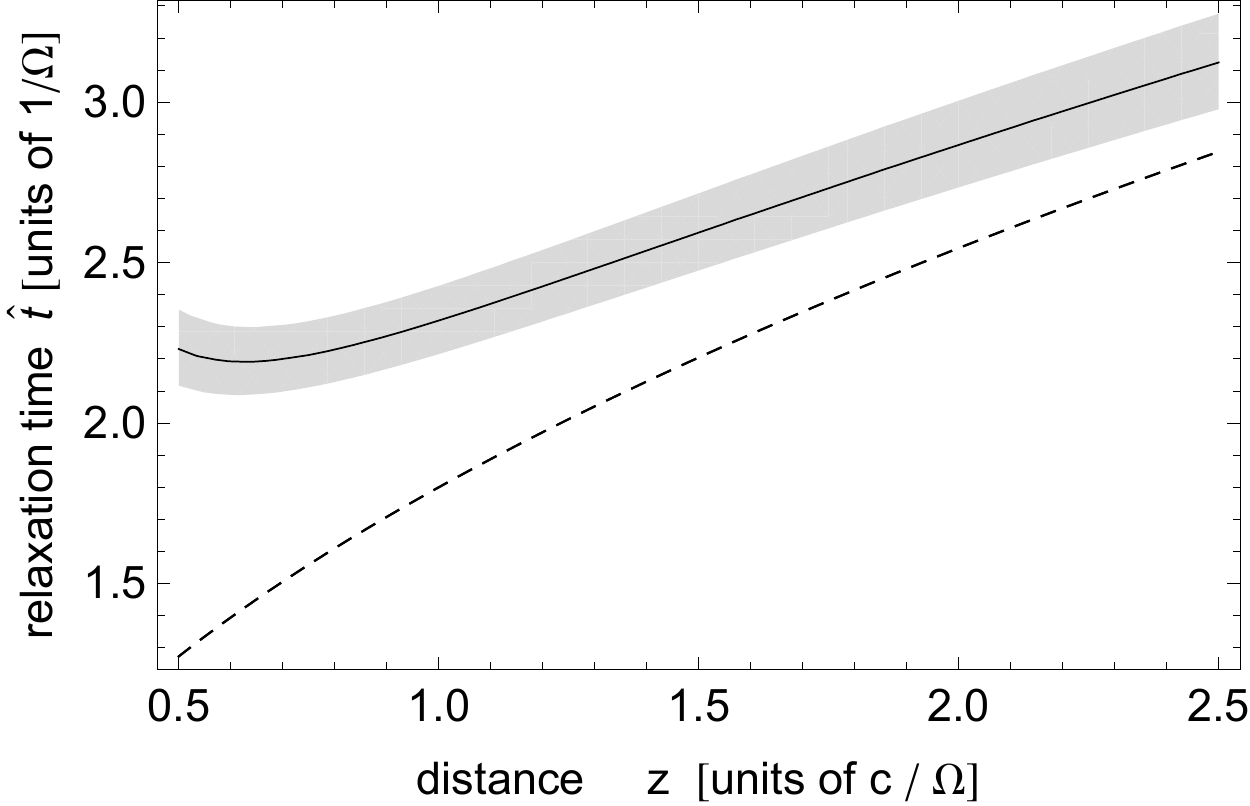}
\caption{%
Time scale for the transient atom-surface dressing vs.\ distance. %}{Surface mode relaxation time vs. distance.} 
Parameters are those of a rubidium atom near a gold surface, as in Fig. \ref{ch_Dyn-fig:dynamic_CP}. 
Shaded band with solid line: numerically calculated $1/e$ decay time of the dynamic Casimir-Polder energy,
with 5\% error margins. Dashed line: estimate $\tau \pi [z / (\omega_{\rm sp} c)]^{1/2}$
derived in the main text, with fitting parameter $\tau = 1.1$.
At short distance the plasmon spectrum becomes monochromatic and losses limit the relaxation time, see Sec. \ref{ch_Dyn-sec:near_field_dressing}.
}
\label{ch_Dyn-fig:decay_time_intermed}
\end{figure}
%================================================

At intermediate distances $c/\omega_{\rm p} < z <  2 c/\Omega$, the time scale of the decay can be estimated from the exponential in the integral in Eq. \eqref{ch_Dyn-eq:pole_contribution4}, the rest of the integrand being a well-behaved function that vanishes as $p\to 0$ and is polynomial at infinity.
From the distance-dependent part of the integrand, only values of $p$ such that $ \bar{\kappa}(p) z < 1$ are relevant, so that $p < \hat p = (z^{-4} + \omega_{\rm sp}^2/c^2 z^2)^{1/4}$.
If the oscillating exponential covers much more than half a period in this range of $p$, both the real and imaginary part of the integral will eventually be small on average.
The typical timescale on which the resonant contribution decays is, therefore, given by $\hat t = \tau \pi  / \bar{\omega}(\hat p) $, with $\tau$ a coefficient of order unity.
For atom-surface separations beyond the plasma wavelength, the time scale can be roughly estimated by
$\hat t \approx \tau \pi [z / (\omega_{\rm sp} c)]^{1/2}$,
the geometric mean between the plasma period and the signal-transit time.

Fig. \ref{ch_Dyn-fig:decay_time_intermed} compares this estimate with a numerical solution.
At short distances, 
large momenta $p$ become relevant and the above 
approximation does no longer apply. The surface spectrum then
strongly peaks at the plasmon resonance.
We will see in the following Section that the relaxation of the dynamic 
Casimir-Polder interaction slows down significant and is limited by 
Ohmic losses in the bulk.

%%%%%%%%%%%%%%%%%%%%%%%%%%%%%%%%%%%%%%%%%%
\subsection{Near-field dressing}
\label{ch_Dyn-sec:near_field_dressing}
%%%%%%%%%%%%%%%%%%%%%%%%%%%%%%%%%%%%%%%%%%
We now consider short distances, $z \ll c/\omega_{\rm p} \ll c/\Omega$, where the surface-mode contribution is dominant. Dissipative properties of the surface are crucial in the near field and we assume the Drude metal of Eq. \eqref{ch_Dyn-eq:Drude_model} allowing for a finite damping rate $\gamma$.
As a result, the surface modes broaden as shown in Fig. \ref{ch_Dyn-fig:spdr}.

At frequencies connected with electric dipole transitions, this near field regime corresponds to atom-surface separations smaller then the normal skin depth of the metallic surface. Here, an asymptote of the reflected electric Green's tensor is obtained from an expansion for large in-plane wavevectors $p\gg \omega_p / c, \omega/c$
\begin{align}
\label{ch_Dyn-eq:nearfield_FDT}
G(z, \omega)  &\approx
\frac{ \vec{d}^2 + d_z^2 }{32 \pi \varepsilon_0 z^3}r_{\rm p}(\omega)
~,
\\
r_{\rm p}(\omega)&\approx \frac{\varepsilon(\omega) - 1}{\varepsilon(\omega)+1}.
\end{align}
The expression can be  analytically continued to complex frequencies and is well-behaved along the whole imaginary axis.
From here and Eq. \eqref{eq:CP_stat} the well-known static van der Waals-Casimir-Polder potential follows immediately. We recall that for an isotropically polarizable atom near a perfectly  reflecting surface we have
$U_{\rm stat} = -|d|^2 / (48 \pi \varepsilon_0 z^3)$, whereas the potential is reduced by a factor 
$\omega_{\rm sp} / ( \Omega + \omega_{\rm sp} ) \approx 0.8$
for the Drude parameters of gold and the atomic transition considered in the 
numerics~\cite{Wylie1985}.

In the near field, retardation can be neglected and the signal-transit time coincides with $t=0$.
To obtain the time dependent potential from Eq. \eqref{ch_Dyn-eq:energy_shift_dyn_im}, we must 
again consider the isolated pole of the near field Green's tensor inside contour III of Fig. \ref{ch_Dyn-fig:contour}, 
given approximately by
\begin{align}
\bar{\omega} (\infty) =  - \omega_{\rm sp} -\imath \frac{\gamma}{2}~,
\end{align}
up to small corrections ${\cal O}( \gamma^2 / \omega_{\rm sp}^2 )$.
This corresponds to the asymptotically flat regime of the surface mode (cf. Fig. \ref{ch_Dyn-fig:spdr}). The relevant residue of $r_{\rm p}$ [Eq.(\ref{ch_Dyn-eq:Residue})] is now
$R_{\rm p}(\bar \omega(\infty), \infty) = 1$
and results in a resonant contribution
\begin{align}
\Delta U_{\rm res}(z,t) = \frac{|d|^2 \omega_{\rm sp}}{48 \pi \varepsilon_0 z^3}  \Re\!\!\left[ \frac{\EXP{\imath (\Omega - \bar\omega(\infty) )t}}{\Omega-\bar\omega(\infty) }\right]\,.
\label{eq:near_field_resonant}
\end{align}
This cancels exactly with the static Casimir-Polder potential for $t \to 0$.
The physical reason is that the static potential, too, originates mainly 
from surface modes at short distance \cite{Annett1986, Zangwill}.
%
%================================================
\begin{figure}[t!]
\centering
 \includegraphics[height=5cm]{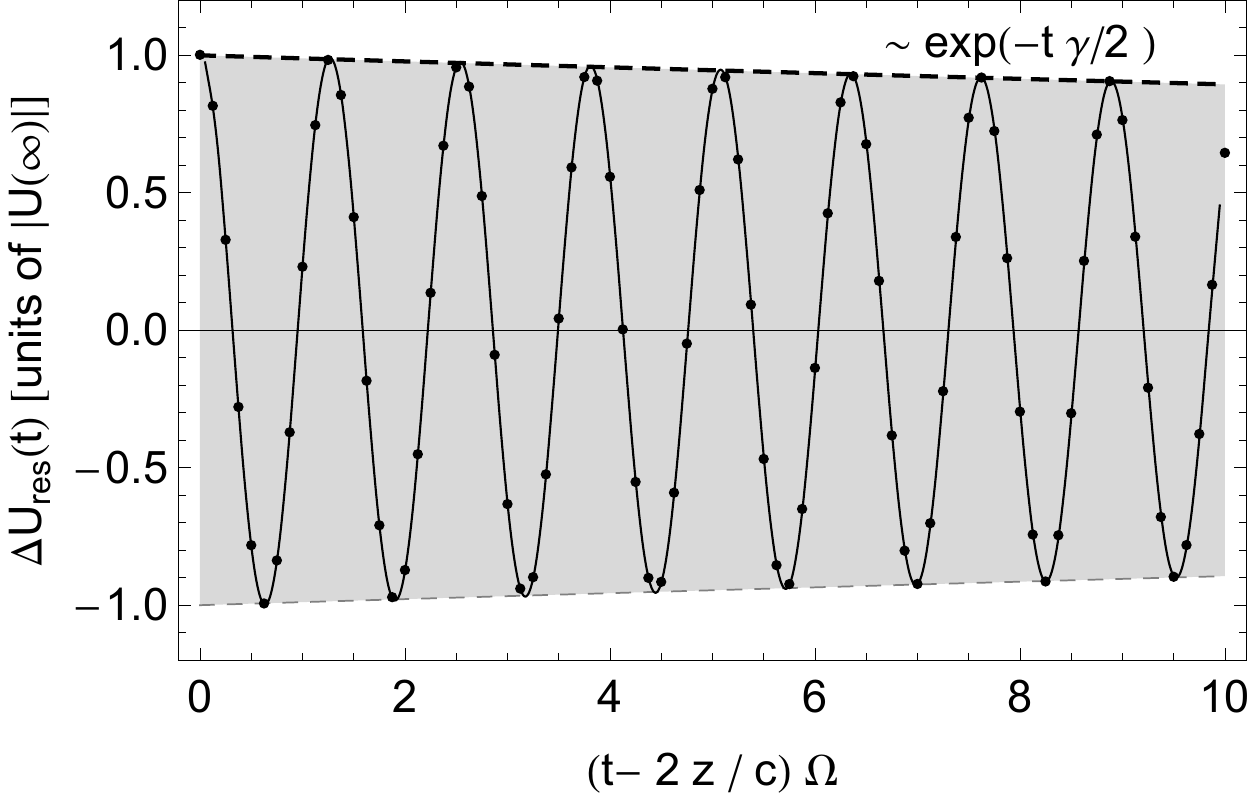}\\[2ex]
\caption{
Dynamic dressing potential after the signal-transit time in the near field ($z = 0.01 c/ \Omega$) for an initially bare atom. The surface is described by a Drude model for gold with parameters as in Fig. \ref{ch_Dyn-fig:dynamic_CP} and the atomic transition corresponds to $780 \unit{nm}$.
The grey envelope shows the exponential damping with a rate $\gamma/2$.
The solid curve corresponds to the numerical evaluation of Eqs. \eqref{ch_Dyn-eq:pole_contribution4} and  \eqref{ch_Dyn-eq:Residue}, and the surface modes for the lossy Drude model. Dots indicate the asymptotic result of Eq. \eqref{eq:near_field_resonant}.
}
\label{ch_Dyn-fig:nearfield_plasmon}
\vspace{1ex}
\end{figure}
%================================================
%
Hence the resonant contribution and the static one are of the same order, as can be seen in the numerical evaluation of the resonant dressing potential at very small atom-surface separations given in Fig. \ref{ch_Dyn-fig:nearfield_plasmon}.  
The plot also demonstrates that the asymptotic results of this section (dots)  
are in excellent agreement with the full calculation based on 
Eq.\eqref{ch_Dyn-eq:pole_contribution4}.
From Eq. \eqref{eq:near_field_resonant}, it is clear that the dynamic
atom-surface potential decays on a time scale set by the spectral width
$\gamma/2$ of the surface plasmon mode.
This is at variance with the result in the far field, where 
the slope of the dispersion relation was the reason for the relaxation.
As another immediate consequence, the resonant potential oscillates at a frequency 
$\Omega + \omega_{\rm sp}$. Again, this differs completely from the far-field result, where the 
oscillation is basically given by the atomic transition frequency
 [see Eq. \eqref{ch_Dyn-eq:pole_contribution4}].

%%%%%%%%%%%%%%%%%%%%%%%%%%%%%%%%%%%%%%%%%%
\section{Partial Dressing After a Nonadiabatic Transition}
\label{sec:partialdressing}
%%%%%%%%%%%%%%%%%%%%%%%%%%%%%%%%%%%%%%%%%%
We now consider the case when the atom-surface system is in a fully relaxed or  
dressed state and some system parameter is suddenly changed. We call this scenario
`partial dressing'.
This has been considered in Ref. \cite{Messina2010} for an atom near a perfect reflector.
Also, the time evolution of the Casimir-Polder force between two neutral atoms starting from a partially dressed state of the system has been investigated in \cite{Passante2003}. 

%%%%%%%%%%%%%%%%%%%%%%%%%%%%%%%%%%%%%%%%%%
%subsection{Partially dressed states}
%%%%%%%%%%%%%%%%%%%%%%%%%%%%%%%%%%%%%%%%%%
We denote $|G\rangle$ the fully-dressed atomic ground state. This state is obtained by perturbation expansion in the basis of the uncoupled Hamiltonian. At first order in the coupling constant
\begin{eqnarray}
|G\rangle 	&=& |\mathrm{vac}, \downarrow\rangle + |\{1\},\uparrow\rangle~,
\\
|\{1\}, \uparrow\rangle		
		&=&  \sum_{|\Psi\rangle \ne |\rm vac,\downarrow\rangle} 
		\frac{\langle\Psi|H_{AF}|\rm vac, \downarrow\rangle}{E_\Psi - E_0}|\Psi\rangle
\nonumber \\
	%	= \dots \nonumber\\
		&=&
		-\imath d_m \int_0^\infty d\omega \int d^3\vec{x} \sqrt{\frac{\varepsilon_0}{\pi\hbar } 
		\Im \varepsilon(\vec x, \omega)}  \nonumber \\
		&& {} \times \frac{\mathcal{G}_{mn}^*(\vec r, \vec x, \omega)}{ \omega +  \Omega} 
		f_{n}^\dagger( \vec{x}, \omega ) | {\rm vac}, \uparrow\rangle~.
		\label{ch_Dyn-eq:interacting_ground_state}
\end{eqnarray}
The sum over possible elements
$\vec{f}^\dagger
|\rm{vac}, \uparrow\rangle$
corresponds to the cloud of virtual bosons surrounding the atom.
Note that \mbox{$\langle G|G\rangle = 1 + \mathcal{O}(d_m d_n)$}  so that normalization corrections result subleading and the dressed state may be considered normalized in the following.

We now suppose that the system is in the stationary state $|G\rangle$ and  that a nonadiabatic (i.e. instantaneous) change in a system parameter occurs at $t=0$.
In the following, the new system observables are marked with a tilde. 
To begin with, we consider
a change of the atomic transition frequency $\Omega \to \tilde \Omega$ which might be induced by the sudden onset of an external electric field via the Stark shift.
This transition does not affect the matter-assisted field part of the system and leaves the structure of the Hilbert space unaltered. The rapidity of the transition makes sure that the quantum state immediately after $t=0$ remains otherwise the same as before.

\begin{figure}[t!]
\centering
\includegraphics[height=5cm]{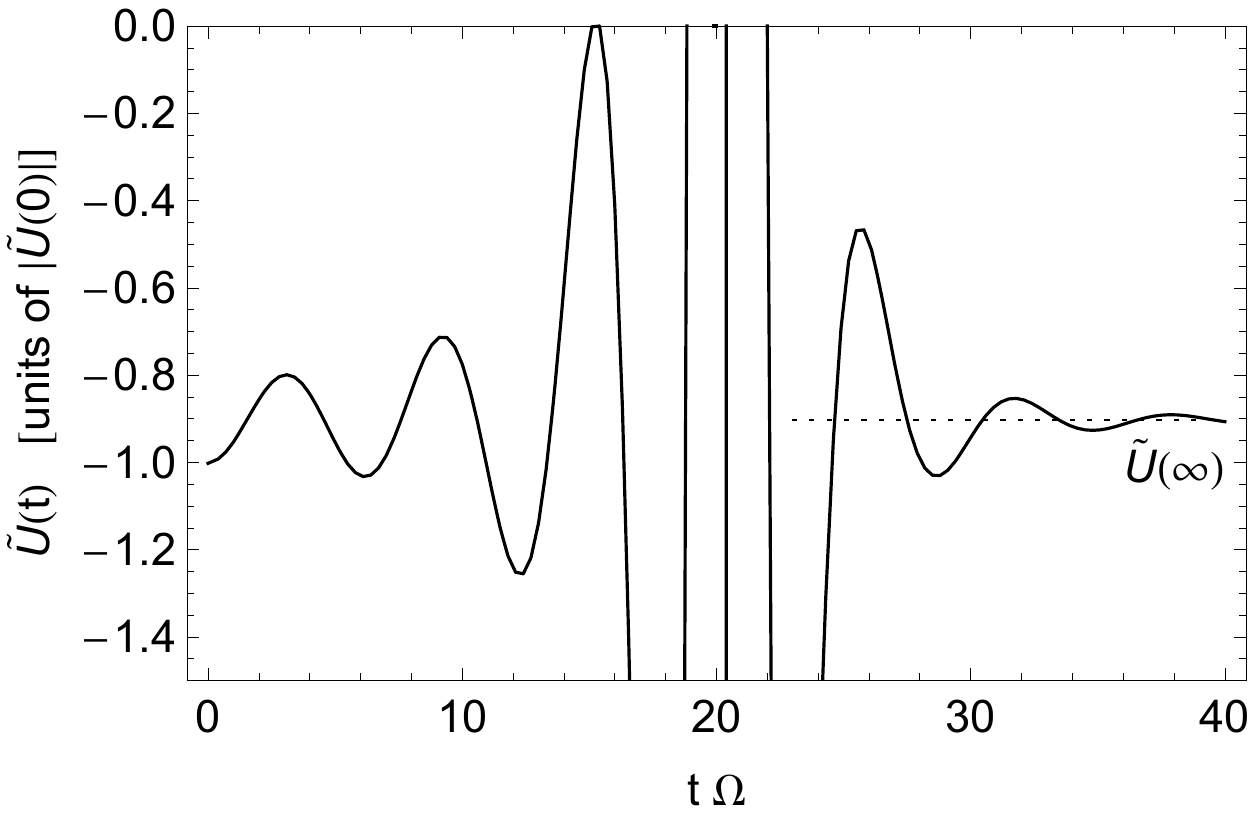}\\[2ex]

\caption{
Partial dressing starting from a dressed ground state, after a sudden change
in the atomic transition $\Omega \to \tilde\Omega$.
The new transition $\tilde\Omega = 0.9\,\Omega$ corresponds a wavelength $780\,\unit{nm}$ and
the surface is described by a Drude model for gold with parameters as in Fig. \ref{ch_Dyn-fig:dynamic_CP}. 
The energy just after the non adiabatic transition, $\tilde U(0)$ is the same as
in the dressed ground state before. 
The divergence occurs around the signal-transit time $2 z / c$. A large distance
$z = 10 \,c/ \Omega$ is considered.
The value of the new static shift $\tilde U(\infty)$ can be understood from the $1/\tilde \Omega$ scaling of the static polarizability \citep{BuhmannBookI}.
}
\label{ch_Dyn-fig:smallstarkshiftFar}
\vspace{1ex}
\end{figure}

The dynamic Casimir-Polder potential
is then calculated in the Heisenberg picture
\begin{eqnarray}
\label{potential_energy}
\tilde U(z, t)  	&=& \textstyle \frac{1}{2}\langle G |\tilde H_{AF}(t)|G\rangle  \\
		&=& \textstyle \frac{1}{2} \langle {\rm vac}, \downarrow |\tilde H_{AF}(t)| { \rm vac}, \downarrow\rangle			\nonumber
\\
		&& {} + \textstyle \frac{1}{2} \left ( \langle {\rm vac}, \downarrow |\tilde H_{AF}(t)|\{1\}, \uparrow\rangle 
		+  \text{c.c.} \right)
		\nonumber
\\
		&& {} + \textstyle \frac{1}{2} \langle\{1\}, \uparrow |\tilde H_{AF}(t)|\{1\}, \uparrow\rangle
		~.	
		\label{ch_Dyn-eq:partial_dressing1}
\end{eqnarray}
Evaluation up to second order in the dipole moment is again sufficient for our purposes. The first term of Eq. \eqref{ch_Dyn-eq:partial_dressing1}  mirrors Eq. \eqref{ch_Dyn-eq:energy_shift_groundstate} and recovers the form of Eq. \eqref{ch_Dyn-eq:energy_shift_complete},
$\tilde U_{\rm stat} + \tilde U_{\rm dyn}$,
with the old transition frequency replaced by the new one. 
The third term is a higher-order contribution and, hence, neglected here. Finally, the second term yields a correction $\Delta \tilde U_{\rm p}$ that is unique to the partial-dressing scenario.
The partial dressing dynamics is hence described by the total time-dependent potential
\begin{align}
\tilde U(z ,t)	= \left[\tilde U_{\rm stat}(z) + \tilde U_{\rm dyn} (z,t)\right] + \Delta \tilde U_{\rm p}(z,t)~,
	\label{ch_Dyn-eq:partial_dressing_complete}
\end{align}
where the partial dressing correction is
\begin{eqnarray}
\Delta \tilde U_{\rm p}(z,t)
	&=&   -\Re \langle \rm{vac}, \downarrow |{\hat{\vec{d}}}^{(0)} \cdot {\vec{E}}^{(0)}|\{1\}, \uparrow\rangle \nonumber\\
	&=&- d_m  d_n \Re \int_0^\infty \frac{d\omega}{2 \pi}
	\frac{2 \EXP{-\imath (\omega + \tilde \Omega) t}}{\omega + \Omega}
	\nonumber\\
	&& \times \int d^3\vec{x}\,
	\varepsilon_0 \, \Im \varepsilon(\vec{x}, \omega) \nonumber\\
	&&\times {\mathcal{G}}_{ml}(\vec{r}, \vec{x}, \omega) \mathcal{G}_{nl}^*(\vec{r}, \vec{x}, \omega)~.
	\label{ch_Dyn-eq:partial_dressing2}
\end{eqnarray}
As before, the spatial integral in Eq. \eqref{ch_Dyn-eq:partial_dressing2} can be performed using Eq.\, \eqref{ch_Dyn-eq:knoells_magical_formula}.
This gives the concise expression
\begin{eqnarray}
	\label{ch_Dyn-eq:partial_dressing_stark_shift}\,
&& \tilde U(z,t) = - % d_m d_n 
\int_0^\infty \frac{d\omega}{2 \pi}\Im [
	G(z, \omega ) 
]
\\
&&\times
\left[\frac{2}{\tilde\Omega + \omega} -  \frac{2 \cos[ (\tilde\Omega + \omega)t ]}{\tilde\Omega + \omega}
+\frac{2 \cos[ (\tilde\Omega + \omega)t ]}{\Omega + \omega}	\right]
\,,
\nonumber
\end{eqnarray}
using again the shorthand $G( z, \omega )$ 
defined in Eq.(\ref{eq:def-shorthand-G}) above.
The fractions in brackets correspond to the three terms in 
Eq.(\ref{ch_Dyn-eq:partial_dressing_complete}).
The mathematical structure shows immediately how this potential 
connects the old stationary state  ($t=0$, provided by the last term) and the new one ($t \to \infty$,  first term). 
This is also clearly visible in Fig. \ref{ch_Dyn-fig:smallstarkshiftFar} which presents the time-dependent potential during the partial dressing. 
In the far-field situation considered here, the surface modes do not contribute strongly and the results resemble those obtained in the case of a perfect reflector \cite{Messina2010}. The latter limit can be recovered analytically from Eqs. \eqref{ch_Dyn-eq:partial_dressing_stark_shift} and \eqref{ch_Dyn-eq:GT_perf}.

%
%================================================
%================================================
 %
%================================================
\begin{figure}[t!]
\centering
a)\includegraphics[height=5cm]{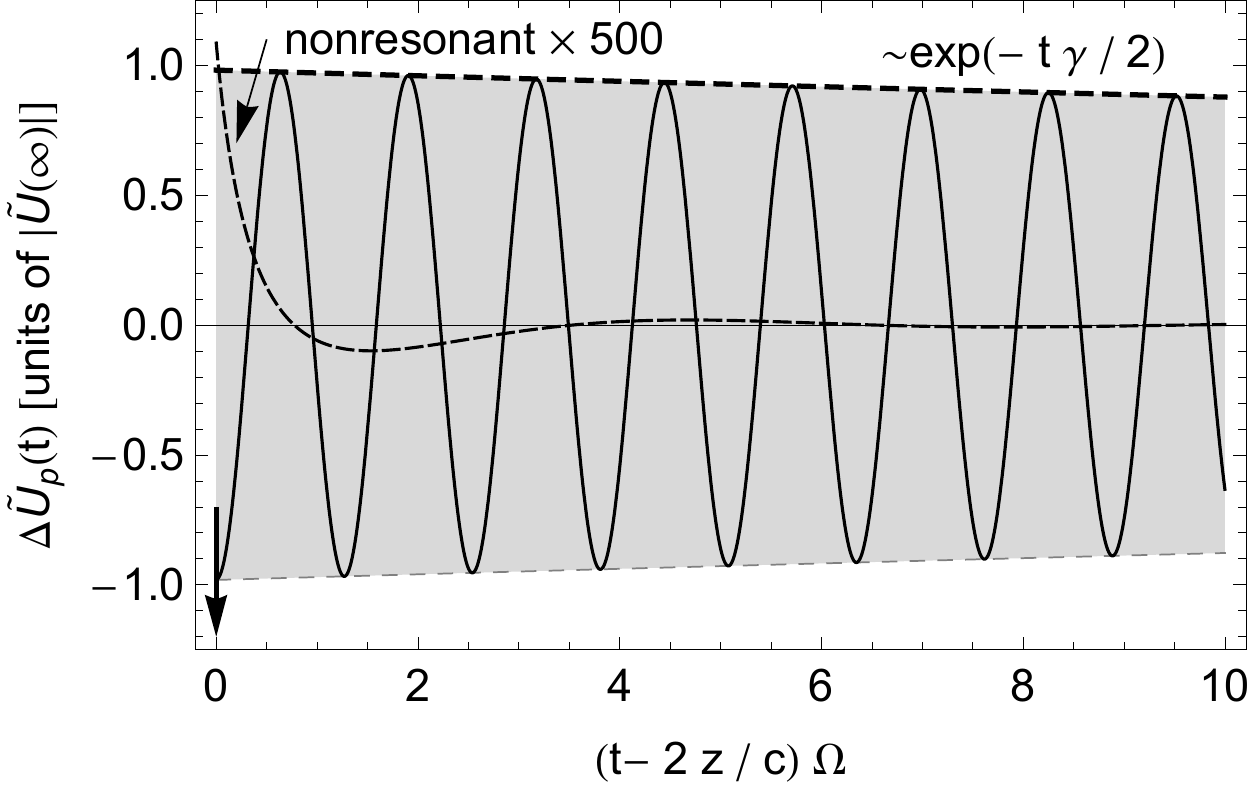}\\[2ex]
b) \includegraphics[height=5cm]{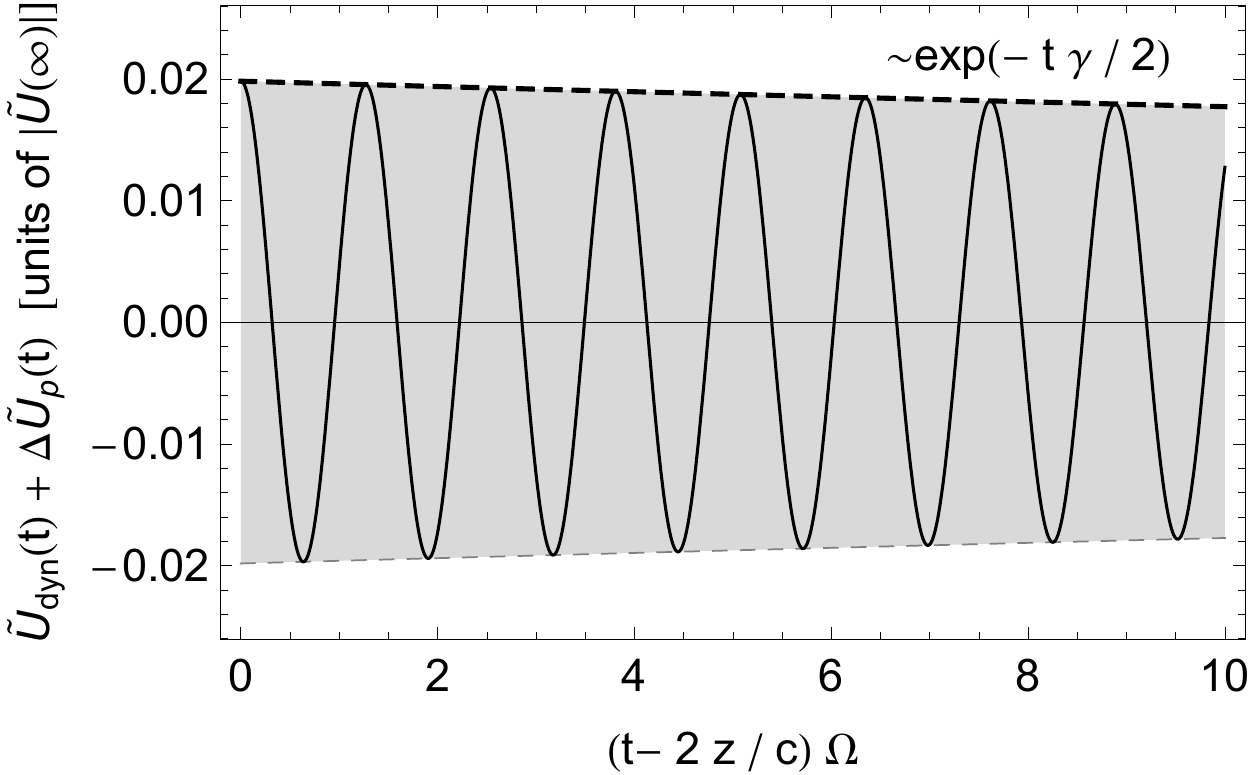}\\[2ex]

\caption{
a)
Partial dressing correction $\Delta\tilde U_{\rm p}(t)$ after the signal-transit time in the near field ($z = 0.01 c/ \Omega$) after the atomic transition is shifted so that the new value $\tilde\Omega = 0.9\Omega$ corresponds to the transition wavelength $780 \unit{nm}$. The surface is described by a Drude model for gold with parameters as in Fig. \ref{ch_Dyn-fig:dynamic_CP}. The grey envelope shows the exponential damping with a rate $\gamma/2$. The dashed curve shows the nonresonant contribution to the partial dressing potential and has been multiplied by a factor 500 for visibility.
b)
Total time-dependent contribution to the partial dressing potential in the same scenario.
At the distance and parameters chosen, the contribution $\tilde U_{\rm dyn} =  U_{\rm dyn}|_{\Omega\to\tilde\Omega}$ coincides with the curve given in Fig. \ref{ch_Dyn-fig:nearfield_plasmon}. According to Eq. \eqref{ch_Dyn-eq:partial_dressing_complete}, the two contributions must be added up, resulting in a reduced amplitude of the plasmonic oscillations.}
\label{ch_Dyn-fig:smallstarkshift}
\vspace{1ex}
\end{figure}
%================================================
%

The partial dressing interaction can be analyzed along the 
lines of Sec.\ref{ch_Dyn-sec:material}, considering separately the plasmon-pole contribution 
(contour III in Fig.\ref{ch_Dyn-fig:contour})
and the nonresonant one [imaginary-frequency integrals in Eq.(\ref{ch_Dyn-eq:energy_shift_dyn_im})].
Fig. \ref{ch_Dyn-fig:smallstarkshift}(a) shows the results for the same Stark shift as in Fig. \ref{ch_Dyn-fig:smallstarkshiftFar}, but at short distance.
Parameters are chosen such that the terms in brackets in Eq. \eqref{ch_Dyn-eq:partial_dressing_complete} coincide with the curve shown in  Fig. \ref{ch_Dyn-fig:nearfield_plasmon}. 
The partial dressing correction $\Delta \tilde U_p$ is comparable in amplitude 
to these terms, but carries the opposite sign.
The total potential after the frequency
shift oscillates around the new static value with a reduced amplitude as shown in
Fig. \ref{ch_Dyn-fig:smallstarkshift}(b).

A closely related scenario involves a change in the transition dipole $\vec{d} \to\tilde {\vec{d}}$.
This can occur, e.g., if an external field induces a sudden (orientation) polarization of the atom. 
The dynamic Casimir-Polder potential is obtained as before and results in a slightly different 
formula where the partial dressing correction depends on both the old and the new dipole moment
\begin{align}
	\nonumber
\tilde U(z,t) &=  \tilde U_{\rm stat}(z)  +  \tilde{d}_m  (\tilde{d}_n - d_n ) 
\\
& {} \times \int_0^\infty \frac{d\omega}{\pi} \Im [\mathcal{G}_{mn}(\omega ) ]
\frac{\cos[ (\Omega + \omega)t ]}{\Omega + \omega}
~.	
	\label{ch_Dyn-eq:partial_dressing_polarization}
\end{align}
%
%\textcolor{red}{\cancel{
%
%In this case the total dynamic dressing potential does not recover the old static value  right after the transition, i.e. $U(\infty) \ne \tilde{U}(0)$. The discontinuity indicates the direct participation of the atom in the nonadiabatic transition and, hence, a polarization energy of the atom. In the extreme case where the old and the new transition dipole are mutually orthogonal (${\vec{d}}\cdot \tilde{\vec{d}}=0$), the partial dressing correction vanishes indicating that the dipole decouples from the cloud of virtual photons by which it was previously dressed. Similarly, a large increase in the dipole moment $|\tilde{\vec{d}}| \gg  |{\vec{d}}|$ (as in the excitation to a highly excited Rydberg state) turns the term involving the old dipole moment negligible.  Clearly, the formal analogy with Eq. \eqref{ch_Dyn-eq:energy_shift_complete} underlines how these particular partial dressing scenarios  realize the bare dressing considered in Sec. \ref{ch_Dyn-sec:bare_atom_dressing}}}.
%%
In this case, there is a discontinuity in the potential at $t=0$. This is indeed expected from a physical point of view, because the interaction energy between the atom and the local field at its position is immediately affected by the change in the atomic dipole.
When ${\vec{d}}\cdot \tilde{\vec{d}} \ll  |\tilde{\vec{d}}|^2$  (by amplitude or orientation), the partial dressing correction is negligible. This agrees with the intuitive notion of ``switching on the interaction'' in the bare atomic dressing of Sec. \ref{ch_Dyn-sec:bare_atom_dressing}.

%==================================
\section{Summary and Discussion}
\label{sec:Summary and Discussion}
%==================================
The dynamic atom-surface Casimir-Polder interaction has been investigated from the perspective of atomic self-dressing for an initially bare or partially dressed atom near a surface described by 
general optical reflection coefficients. In the more realistic scenario of a partially dressed state, the system relaxes after a sudden change in the atomic transition frequency or the transition dipole. We have thus generalized previous work on dynamic Casimir-Polder interaction \cite{Vasile2008, Messina2010} with perfectly reflecting surfaces to more realistic ones using the 
quantization scheme for matter-assisted fields to include boundaries with, in principle, arbitrary dielectric or magnetic properties.
Although the limit of perfectly reflecting surfaces can be often considered as a reasonable approximation in the radiative zone, that is for distances far beyond the atomic transition wavelengths, many interesting phenomena require a more detailed surface model, in particular when shorter distances are considered. This is not unexpected, because surface excitations are known to dominate the response in the near-field zone \cite{Ford1984, Henkel2002, Failache2002, Gorza2006,  Stehle2011}.

Within this framework, we have studied the dynamic dressing and the consequent dynamic Casimir-Polder potential between a two-level atom and a surface in different nonequilibrium situations: an initially bare ground-state atom and a partially dressed atom.
We have  compared the results obtained using parameters typical of a gold surface (Drude model) with recent results in the literature obtained for a perfectly reflecting surface and isolated the resonant surface-mode contributions to the time-dependent atom-surface potential in different distance regimes. Far from the surface, this contribution is a minor correction to the dynamic atom-surface potential and decays quickly ($\propto t^{-5}$) due to the dephasing of different frequency components.
In contrast, at short distances the dynamic Casimir-Polder interaction bears the characteristics of a transient excitation of the surface plasmon modes that dominate the response by orders of magnitude and decays on a much slower scale related to the electron scattering time in the bulk metal.
Both for the initially bare atom and for the partially dressed one, we have found that time intervals exist where the Casimir-Polder potential is repulsive. 
The particular case of an atom dressing after a change in its transition dipole moment may effectively provide a realization of the bare dressing scenario.

Our formalism may be immediately generalized to other types of emitters such as small molecules, quantum dots, Rydberg states, or -- in analogy to the dynamic Casimir effect studied in Ref. \cite{Johansson2009} -- superconducting flux qubits.
%
%All of these have slower dynamic time scales than the ground state atoms considered in this work. 
Among these, the latter two feature a considerable magnetic polarizability, leading to a high sensitivity towards the low-frequency electromagnetic modes \cite{Haakh2009b, Intravaia2009, Intravaia2010}. 
A study of such systems would contribute to the understanding of the role of the DC conductivity in the macroscopic Casimir interaction \cite{Bimonte2005, Bimonte2010}.
A particularly interesting scenario could be the partial dressing triggered by a phase transition in the surface which quickly changes the optical properties, e.g. in the evolution of a magnetic dipole close to a superconductor near criticality.
Knowledge of the atomic response to a perturbation of superconductivity is also relevant for experiments with superconducting atom chips \cite{Reichel2011}.
In this case, the formalism of Sec. \ref{sec:partialdressing} must be amended to ensure a fully retarded signal at the dipole position.
%, which takes place on a time scale of $1\dots100\unit{ns}$ \cite{Geier1982} comparable to the dynamic time scales of magnetic dipoles.
%In fact, the 

%
%
%Further scenarios in which partial dressing could be observed include a phase transition in the surface that quickly changes the optical properties. In these cases, the transition directly affects the coupling Hamiltonian and the previous treatment must be generalized carefully.
%A system that seems very promising for both the dynamic dressing and the dynamic Casimir effect is a superconductor, in which a phase transition can be induced, e.g. by a supercritical current-pulse or by a strong laser pulse.
%Superconductivity breaks down on the scale of $1\sim 100\,\unit{ns}$  \cite{Geier1982}, rather slow on the scale of the electronic dynamics of ground-state atoms, but comparable to the relevant dynamical scales in  small molecules, quantum dots, Rydberg states, or -- in analogy to the dynamic Casimir effect studied in Ref. \cite{Johansson2009} -- superconducting flux qubits.
%Among these, the latter two feature a considerable magnetic polarizability. 
%This leads to a high sensitivity towards the low-frequency electromagnetic modes \cite{Haakh2009b, Intravaia2009, Intravaia2010}. 
%Such a study would also contribute to the understanding of the role of the DC conductivity in the macroscopic Casimir interaction \cite{Bimonte2005, Bimonte2010}.
%Knowledge of the atomic response to a perturbation of superconductivity is also relevant for experiments with superconducting atom chips \cite{Reichel2011}.

Finally, the thermal occupation of the field and surface modes \cite{Ribeiro2013} may affect the dynamic dressing. 
Thermal excitations can be absorbed by the atom, giving rise to additional resonant contributions.  
The width of the thermal spectrum is likely to introduce a characteristic time scale $\hbar / k_B T \approx 1\,\unit{ps} \times (T / 1 \unit{K})^{-1}$. 
If thermal occupation of excited atomic states is relevant, 
spontaneous emission will play a role in the dynamic Casimir-Polder
interaction.
The study of dressing may thus  give some insight into the process of thermalization.

\begin{acknowledgments}
Partial financial support by Deutscher Akademischer Austauschdienst (grant D/10/41486), the German-Israeli Foundation for Scientific Research and Development, by the Julian Schwinger Foundation, by MIUR, and by Comitato Regionale di Ricerche Nucleari e di Struttura della Materia  is gratefully acknowledged.
The  authors also acknowledge support from the ESF Research Networking Program CASIMIR and thank R. Messina and F. Intravaia for helpful discussions.
\end{acknowledgments}

%%%%%%%%%%%%%%%%%%%%%%%%%%%%%%%%%%%%%%%%%%%%%%%%%%%%%%
\appendix
\bigskip
\section{Electric Green's tensor} \label{ch_App_em}

%%%%%%%%%%%%%%%%%%%%%%%%%%%%%%%%%%%%%%%%%%%%%%%%%%%%%%
The retarded electric Green's tensor describes the electric field
\begin{equation}
E_{i}(\vec{x}, \omega ) = \mathcal{G}_{ij}( \vec{x}, \vec{r}, \omega )
d_j(\omega)	%~,\Quay
%
%B_{i}(\vec{r}, \omega ) = \mathcal{H}_{ij}( \vec{r}, \vec{r}', \omega )
%m_j(\omega)	~,
	\label{ch_App_em-eq:def-Hij}
\end{equation}
radiated by a point-like fictitious dipole source $\vec{d}$ placed in $\vec{r}$.
% \cite{note2}.
Note that other authors use a different normalization for the Green tensor; 
in Refs. \cite{Knoll2001, Safari2006,Buhmann2007,Scheel2008, BuhmannBookI}, 
a prefactor $\omega^2/(\varepsilon_0 c^2)$ would appear in 
Eq. (\ref{ch_App_em-eq:def-Hij}).

Near a single surface, this field consists of a free-space part and a reflected contribution
$%\begin{align}
\boldsymbol{ \mathcal{G}} = \boldsymbol{ \mathcal{G}}^{\rm free} +\boldsymbol{ \mathcal{G}}^{\rm refl}.
$ %\end{align}
We concentrate in this work on the reflected field contributions, that dominate the imaginary part of the Green's tensor by far
at short distances. Neglecting the free space contribution also provides a convenient way of removing the formally diverging Lamb shift from the interaction potentials \cite{Wylie1985}.

The general expression for the reflected (retarded) Green's tensor can be conveniently expressed in the Weyl basis of transverse s- and p-polarized waves, also known as TE- and TM-polarization, respectively \cite{Sipe1981}.
For  a local, isotropic, nonmagnetic bulk medium,
the reflection coefficients read
\begin{equation}
r_{\rm s}(\omega, p)=\frac{\kappa-\kappa_{m}}{\kappa+\kappa_{m}},
\quad
r_{\rm p}(\omega, p)=\frac{\varepsilon(\omega)\kappa-\kappa_{m}}{\varepsilon(\omega)\kappa+\kappa_{m}}~,
\label{Fresnelcoeff}
\end{equation}
where $\kappa = \sqrt{p^2 - \omega^2/c^2}$ and $\kappa_{m} = \sqrt{p^2 - \varepsilon(\omega)\omega^2/c^2}$
 ($\!\Re \kappa \ge 0$, $\Im \kappa \le 0$ for $\omega>0$)
are the propagation constants in vacuum and inside the bulk.
In the limit $\vec{x}  \to \vec{r} $, we obtain the single-point reflected Green's tensor, which depends only on the distance $z$ from the surface (SI units)
\begin{widetext}
\begin{eqnarray}
\mathcal{G}_{ij}^{\rm refl} (z, \omega)= \frac{1}{8 \pi \varepsilon_0} \int_0^\infty \frac{ p~� dp}{\kappa}\, e^{-2 \kappa z}
	\biggl[
				\left(\kappa^2 r_{\rm p}(\omega, p)
				+ \frac{\omega^2}{c^2} r_{\rm s}(\omega, p) \right)
				 [\delta_{ij} -  \hat{z}_i \hat{z}_j]  
				 +2 p^2 r_{\rm p}(\omega, p) \hat{z}_i \hat{z}_j
	\biggr]~.
\label{ch_App_em-eq:mag_greentensor}
\end{eqnarray}
\end{widetext}
From this expression, the limit of an ideally reflecting surface of Eq. \eqref{ch_Dyn-eq:GT_perf} can be obtained  by setting \mbox{$r_{\rm p} = - r_{\rm s} \to  1$}.
For an isotropic polarizability, only the
trace of the tensor is required.

Note finally, that at very short distances from the surface, high values of $p\gg \omega/c$ dominate the integral so that $\kappa, \kappa_m \approx p$. Here, the reflection coefficients are approximated by
%\begin{align}
$r_{\rm p} \approx [\varepsilon(\omega) - 1] / [\varepsilon(\omega) + 1]$, 
$r_{\rm s} \approx (\varepsilon(\omega) - 1) \omega^2 / (4 c^2 p^2)$.
%\end{align}
The s-polarized contribution to the Green's tensor is suppressed by a factor $\omega^2/(c p)^2$ and can be neglected. This gives the near field limit  of Eq. \eqref{ch_Dyn-eq:nearfield_FDT}.
%

%%%%%%%%%%%%%%%%%%%%%%%%%%%%%%%%%%%%%%%%%%%%%%%%%%%%%%

\end{document}